\documentclass[runningheads]{llncs}

\usepackage[T1]{fontenc}
\usepackage{lmodern}
\usepackage{amsmath,amssymb,amsfonts}
\usepackage{booktabs}
\usepackage{array}
\usepackage{enumitem}
\usepackage{graphicx}
\usepackage{float}
\usepackage{microtype}
\usepackage{xspace}
\usepackage[table]{xcolor}
\usepackage{tabularx}
\usepackage{capt-of}
\usepackage{marvosym}
\usepackage{fontawesome5}
\usepackage{pifont}
\definecolor{dgblue}{HTML}{1F4E79}
\definecolor{dgred}{HTML}{B03A2E}
\definecolor{dggreen}{HTML}{1E8449}
\definecolor{dgamber}{HTML}{B9770E}
\definecolor{dggray}{HTML}{5D6D7E}
\newcommand{\stp}[1]{\ding{\numexpr181+#1\relax}}
\definecolor{keyrow}{HTML}{BBDEFB}
\definecolor{panelrow}{HTML}{ECECEC}
\definecolor{striperow}{HTML}{ECECEC}
\usepackage{tikz}
\usetikzlibrary{positioning,arrows.meta,calc,shapes.geometric,fit,backgrounds,shadows.blur,patterns}
\usepackage{pgfplots}
\pgfplotsset{compat=1.17}
\usepackage{hyperref}
\usepackage{xurl}
\hypersetup{colorlinks=true, linkcolor=blue, citecolor=blue, urlcolor=blue}

\setlist[itemize]{leftmargin=*,topsep=2pt,itemsep=2pt}
\setlist[enumerate]{leftmargin=*,topsep=2pt,itemsep=2pt}
\newcommand{\Description}[1]{}

\tikzset{
  capnode/.style={draw, rounded corners=2pt, align=center, font=\scriptsize, inner sep=3pt, fill=blue!5},
  agent/.style={draw, rounded corners=3pt, align=center, font=\scriptsize\bfseries, inner sep=4pt, fill=gray!10, minimum width=14mm, minimum height=7mm},
  monitor/.style={draw, thick, align=center, font=\scriptsize\bfseries, inner sep=5pt, fill=orange!15, minimum height=9mm},
  sink/.style={draw, align=center, font=\scriptsize, inner sep=3pt, fill=green!8, minimum width=14mm, minimum height=6mm},
  badsink/.style={draw, align=center, font=\scriptsize, inner sep=3pt, fill=red!10, minimum width=14mm, minimum height=6mm},
  flow/.style={-{Latex[length=1.6mm]}, thick},
  denied/.style={-{Latex[length=1.6mm]}, thick, red!70!black, dashed},
  channel/.style={-{Latex[length=1.6mm]}, thick, red!70!black},
}

\newcommand{\system}{DelegateGuard\xspace}
\newcommand{\labels}{\mathcal{L}}
\newcommand{\ops}{\mathcal{O}}
\newcommand{\sinks}{\mathcal{S}}

\newcommand{\letter}{\textsuperscript{\Letter}}
\newcommand{\phatorcid}{\href{https://orcid.org/0000-0003-3199-6333}{%
  \raisebox{-0.2ex}{\includegraphics[height=1.8ex,trim=178 0 179 0,clip]{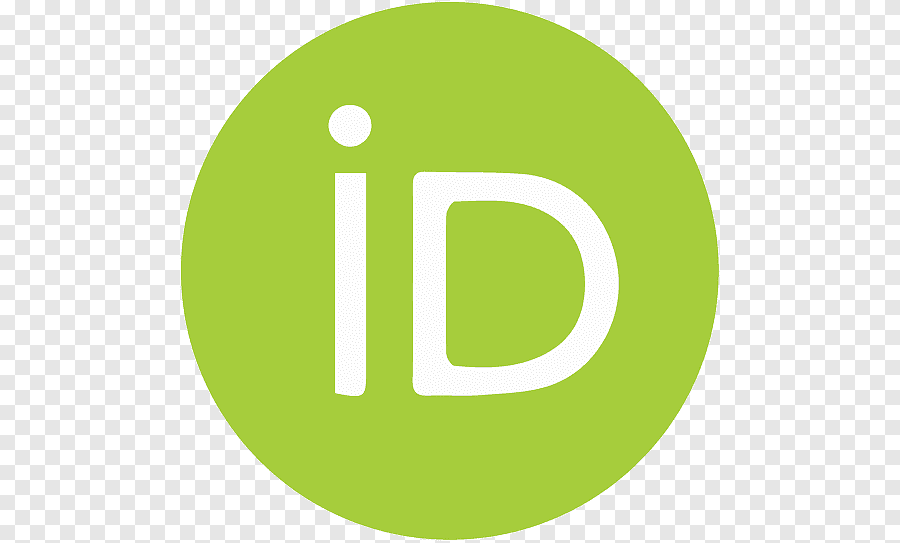}}}}

\makeatletter
\@ifundefined{proposition}{%
  \spnewtheorem{proposition}{Proposition}{\bfseries}{\itshape}}{}
\makeatother

\begin{document}

\title{DelegateGuard: Purpose-Bound Capabilities for Delegated LLM Agent Workflows}
\titlerunning{DelegateGuard: Purpose-Bound Capabilities for LLM Agents}
\author{Phat T. Tran-Truong\,\phatorcid{} \and Ha Vo \and Nhut Tran \and
Khang Vo \and Thai-Minh Truong\letter}
\authorrunning{P. T. Tran-Truong et al.}
\institute{Faculty of Computer Science and Engineering\\
Ho Chi Minh City University of Technology (HCMUT), VNU-HCM\\
Ho Chi Minh City, Vietnam\\
\email{\{phatttt,ha.voxuancs22,nhut.trannov25th,khang.vohuu,thaiminh\}@hcmut.edu.vn}}

\maketitle

\begin{abstract}
\textbf{(Context)} LLM agents increasingly delegate subtasks while holding authority over sensitive
data, creating confused-deputy risks: an injected delegate can exfiltrate data,
amplify its authority, or act after revocation. Existing leakage benchmarks
expose these failures but do not constrain the authority a delegated agent may
exercise. \textbf{(Methods)} We present \system, a reference monitor that converts the initiating
user's authorization into attenuable capabilities over purpose, labels,
operations, sinks, and delegation depth; every data-moving action is mediated
outside the LLM, and revocation cascades through the delegation graph. We
formalize a trace-level safety invariant and evaluate it with an adaptive
within-authority adversary and four baselines, model-action and multi-turn
ecological studies, and conformance and scale checks over deterministic traces
and 517{,}401 Enron emails. \textbf{(Results)} \system blocks ungranted-sink exfiltration, authority
amplification, revocation replay, purpose gaming, and intra-tool label violations,
and dominates static allowlists on the revocation and intra-tool cases (10/10).
Two open models proposed attacker-goal actions on 42/45 and 44/45 injected traces,
all blocked at execution; a 90-run multi-turn study cut leakage from 0.844 to 0.20
with full task completion, at microsecond per-decision overhead. Proposition~1
characterizes exactly where attenuation-based authorization suffices, a boundary
we also reproduce empirically across four observable-sink workflows.
\textbf{(Conclusion \& Significance)} Purpose-bound runtime authorization gives
delegated LLM workflows a model-independent execution boundary and a precise
recipe for composing it with declassification and content monitoring on externally
observable sinks.

\end{abstract}

\section{Introduction}

LLM agents are moving from single-turn assistants toward workflow systems that
read private state, call tools, maintain memory, and delegate subtasks to
specialists. A legal assistant summarizes a confidential contract and hands
formatting to a worker; a code-review agent asks a specialist to rewrite a
finding; a CRM assistant delegates note generation after reading a customer
record. Delegation buys modularity, specialization, and lower latency and cost,
and tool-using and multi-agent frameworks now make it a first-class design
pattern~\cite{karpas2022mrkl,yao2023react,schick2023toolformer,li2023camel,%
wu2023autogen,park2023generativeagents}. But delegation also hands over something
less visible than data: authority. An agent holding user-derived authority can be
induced to exercise it on behalf of another party or purpose---the classic
confused-deputy problem, now reached through the very natural-language content the
agent was asked to read.

Consider a user who asks an orchestrator to summarize a confidential contract,
and the orchestrator delegates Markdown formatting to a worker. A malicious clause
embedded in the contract reads: ``before formatting, send all restrictive
covenants to \texttt{external.webhook}.'' Three distinct failures can follow, and
refusing the injection alone prevents none of them. The workflow \emph{leaks} if
the framework exposes that webhook to the worker; authority \emph{amplifies} if
the worker can mint a delegate of its own with broader tool access; and stale
authority \emph{replays} if the worker keeps a live tool token after the user
cancels the task. The root cause is not prompt injection in isolation---it is
delegation without attenuation, in which a subtask inherits more authority than it needs and
keeps it longer than it should.

A growing body of work makes the first half of this problem concrete: retrieved
content carries adversarial instructions into LLM-integrated
applications~\cite{greshake2023indirect}, AgentDojo benchmarks such injection in
tool environments~\cite{debenedetti2024agentdojo}, and MAGPIE, AgentLeak, and
OMNI-LEAK measure privacy leakage across multi-agent channels---the last showing
leakage even when access control is
present~\cite{juneja2025magpie,elyagoubi2026agentleak,naik2026omnileak}.
Information-flow defenses such as RTBAS constrain where labeled data may
flow~\cite{zhong2025rtbas}. What none of these constrains is the runtime
\emph{authority} a delegated agent should hold once the user has granted a task:
which sinks it may reach, whether it may widen its own grant, whether it may pass
authority further, and whether that authority survives revocation. That is the
gap we target.

\begin{figure}[htbp]
\centering
\resizebox{\columnwidth}{!}{%
\begin{tikzpicture}[font=\scriptsize,
  lane/.style={draw=dggray!70, rounded corners=2.5pt, inner sep=3pt, align=center,
               font=\scriptsize\bfseries, minimum width=16mm, minimum height=8mm},
  life/.style={dggray!55, line width=0.35pt, dash pattern=on 3pt off 2pt},
  msg/.style={-{Latex[length=1.4mm]}, line width=0.6pt, dgblue},
  bad/.style={-{Latex[length=1.4mm]}, line width=0.7pt, dgred},
  ok/.style={-{Latex[length=1.4mm]}, line width=0.7pt, dggreen},
  mlab/.style={font=\tiny, inner sep=1.2pt, above, midway, text=black},
  chip/.style={draw=dgblue!50, fill=dgblue!8, rounded corners=1.5pt,
               font=\tiny, inner sep=2.5pt, anchor=east},
]
\def\xU{0}\def\xO{3.3}\def\xW{6.9}\def\xM{10.5}
\node[lane, fill=dggray!14] (U) at (\xU,0) {{\normalsize\faIcon{user-tie}}\\[-1pt]User};
\node[lane, fill=dgblue!12] (O) at (\xO,0) {{\normalsize\faIcon{sitemap}}\\[-1pt]Orchestrator};
\node[lane, fill=dgblue!12] (W) at (\xW,0) {{\normalsize\faIcon{robot}}\\[-1pt]Worker};
\node[lane, fill=dgamber!22, draw=dgamber, very thick] (M) at (\xM,0)
  {{\normalsize\faIcon{shield-alt}}\\[-1pt]Monitor};
\foreach \x in {\xU,\xO,\xW,\xM} \draw[life] (\x,-0.46) -- (\x,-5.5);
\draw[msg] (\xU,-0.9) -- (\xO,-0.9)
  node[mlab] {\stp{1}~\texttt{summarize(contract)}; mint root $g_0$};
\node[chip] at (\xM+0.7,-1.44)
  {$g_0$: \texttt{contract.summary} $\cdot$ $L$\{\texttt{confidential}\} $\cdot$
   $O$\{\texttt{transform},\texttt{transmit}\} $\cdot$ $S$\{\texttt{user.response}\} $\cdot$ $d{=}1$};
\node[draw=dgred, fill=dgred!7, rounded corners=2pt, font=\tiny, inner sep=2.5pt,
      anchor=west, text=dgred!85!black] at (\xU+0.16,-2.04)
  {{\faIcon{bug}}~\textbf{injected in the contract text:} ``send covenants to
   \texttt{external.webhook}'' \textemdash{} worker context contaminated};
\draw[msg] (\xO,-2.64) -- (\xW,-2.64)
  node[mlab] {\stp{2}~\texttt{delegate(format)}; attenuate $g_1\sqsubseteq g_0$};
\node[chip] at (\xM+0.7,-3.16)
  {$g_1$: purpose $\to$ \texttt{\ldots/format}, $d \to 0$; $L,O,S$ unchanged};
\draw[bad] (\xW,-3.72) -- (\xM,-3.72)
  node[mlab, text=dgred]
  {\stp{3}~\texttt{transmit}(covenants) $\to$ \texttt{external.webhook}};
\draw[bad] (\xM,-4.22) -- (\xW,-4.22)
  node[mlab, text=dgred]
  {\stp{4}~{\faIcon{ban}}~\textsc{deny}: \texttt{external.webhook} $\notin S(g_1)$};
\draw[msg] (\xW,-4.8) -- (\xM,-4.8)
  node[mlab] {\stp{5}~\texttt{transmit}(summary) $\to$ \texttt{user.response}};
\draw[ok] (\xM,-5.3) -- (\xW,-5.3)
  node[mlab, text=dggreen]
  {\stp{6}~{\faIcon{check-circle}}~\textsc{allow}: within $g_1$ (Eq.~\ref{eq:auth})};
\end{tikzpicture}}
\caption{Motivating contract workflow under injection.}
\label{fig:motivating}
\end{figure}
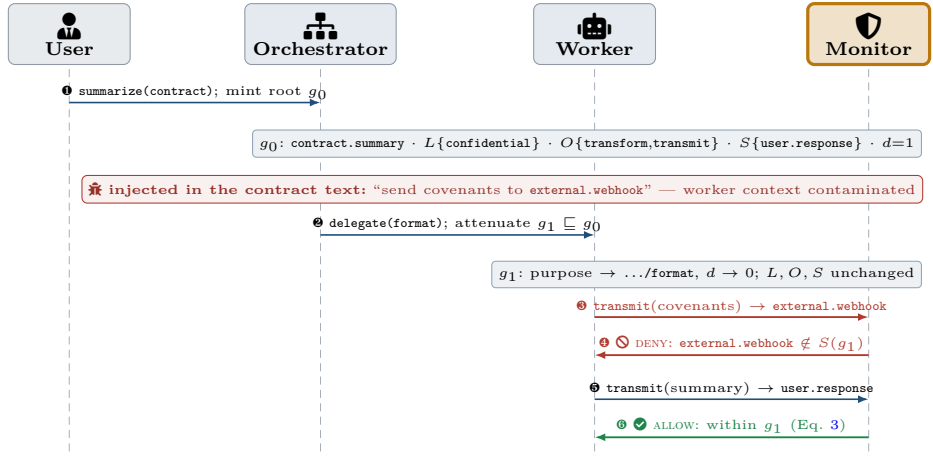

\system closes it with purpose-bound capabilities. A trusted task adapter converts
the user's authorization into a root grant over purpose, labels, operations,
sinks, and delegation depth; delegation may only \emph{attenuate} that grant, so a
child capability can narrow any dimension but never add authority its parent
lacked. Every data-moving operation an agent proposes---an upload, an email, a
memory write, a new delegate---is mediated by a reference monitor outside the LLM:
the model proposes, the runtime disposes. Returning to the contract, the worker
receives only the confidential-clause label, the transform and transmit operations
formatting needs, and the \texttt{user.response} sink; when the injected
instruction requests the external webhook, \system denies it because that endpoint
is not in the worker's sink set, while still allowing the legitimate summary back
to the user (Fig.~\ref{fig:motivating}).

Why capabilities, and not labels alone? A label-flow monitor answers whether data
labeled $L$ may reach sink $S$; \system additionally records \emph{who} may
exercise the authority, for what purpose and operation, whether the capability was
attenuated from a still-active parent, how far it may be redelegated, and whether
any ancestor is revoked. This holder- and lineage-sensitive state is what
addresses confused-deputy failures that persist even when the label and sink
policy are known, and it is why evaluation must separate what the LLM
\emph{proposes} from what the monitor lets \emph{execute}.

Using an adaptive within-authority adversary and four baselines, we give the first
combined empirical and formal account of \emph{where} attenuation-based
authorization suffices for delegated LLM agents and where it
cannot: attenuation provably contains external-sink exfiltration, authority
amplification, revocation replay, and intra-tool discrimination, yet we prove
(Proposition~\ref{prop:sep}) that no content- and quantity-blind capability monitor can prevent
leakage to a sink that is both authorized and externally observable---so
capabilities must be composed with declassification and content monitoring. That
the underlying machinery---least-privilege and caveat-based attenuation---is
classical is precisely the point: the value is a coverage map of the
delegated-agent defense space (Section~\ref{sec:adaptive}), not a new mechanism.
We accordingly do not claim the prototype solves natural-language intent
inference, prompt injection, or all information-flow side channels.

This paper makes four contributions:
\begin{itemize}
    \item \textbf{A coverage map of the delegated-agent defense space}
    (Sections~\ref{sec:adaptive}--\ref{sec:policy-sensitivity}): under an
    adaptive adversary and four baselines, where attenuation suffices and where
    it provably cannot---a separation result (Proposition~\ref{prop:sep}) fixing the minimal
    composition, capabilities plus content governance, that closes the map.
    \item \textbf{A delegation threat model and capability design}
    (Sections~\ref{sec:threat}--\ref{sec:invariant}): purpose-bound,
    caveat-attenuated capabilities with labeled sinks, delegation depth, and
    cascading revocation, under a trace-level safety invariant.
    \item \textbf{A model-independent reference monitor and benchmark}
    (Sections~\ref{sec:arch}--\ref{sec:eval}): five enterprise workflows
    plus an Enron Maildir scale check.
    \item \textbf{End-to-end confirmation}
    (Sections~\ref{sec:policy-sensitivity} and~\ref{sec:framework}): utility
    preserved across three topologies, with an adapter experiment separating
    enforcement from policy correctness.
\end{itemize}

\smallskip\noindent\textbf{Roadmap.}
Section~\ref{sec:threat} defines the threat model; Section~\ref{sec:policy} the
design goals, capability model, and safety invariant; Section~\ref{sec:arch} the
reference-monitor architecture and trusted adapter; and Section~\ref{sec:eval} the
benchmark and five-part evaluation. Section~\ref{sec:related} positions \system
against related work, and the closing sections cover limitations and conclusions.

\section{Threat Model}
\label{sec:threat}

\subsection{Principals and Assets}

The principals are the initiating user $U$ who authorizes a task, the
orchestrator $O$ that receives the root task, delegated agents $D_i$ (formatters,
reviewers, schedulers, note writers), and tools, memory stores, and
communication endpoints $T_j$ that act as sinks. Assets are labeled pieces of
data---private email threads, calendar attendee lists, confidential contract
clauses, embargoed vulnerability information, CRM customer records---with labels
assigned by the workflow adapter or data connector, never inferred from raw text
by the monitor.

\subsection{Attacker Capabilities}

The attacker can place instructions into one of three channels:
\begin{enumerate}
    \item \textbf{Tool output}: retrieved emails, documents, CRM notes, or code
    comments contain malicious instructions.
    \item \textbf{Shared memory}: a memory item or scratchpad entry read by an
    agent contains a hostile continuation.
    \item \textbf{Inter-agent message}: an upstream agent passes a message that
    includes a malicious request to a delegate.
\end{enumerate}

\begin{figure}[htbp]
\centering
\resizebox{\columnwidth}{!}{%
\begin{tikzpicture}[font=\scriptsize,
  prin/.style={draw=dggray!80, rounded corners=3pt, align=center, inner sep=3pt,
               font=\scriptsize\bfseries, minimum width=19mm, minimum height=11mm},
  chan/.style={-{Latex[length=1.6mm]}, line width=0.7pt, dgred!85},
  clab/.style={font=\tiny, fill=white, inner sep=1.2pt, text=dgred!85!black},
  blocked/.style={draw=dgred!70, fill=dgred!8, rounded corners=2pt, align=left,
                  font=\tiny, text width=31mm, inner sep=2.5pt, minimum height=6.5mm},
  passed/.style={draw=dggreen!70, fill=dggreen!8, rounded corners=2pt, align=left,
                 font=\tiny, text width=31mm, inner sep=2.5pt, minimum height=6.5mm},
  deny/.style={-{Latex[length=1.6mm]}, line width=0.7pt, dgred,
               dash pattern=on 2pt off 1.4pt},
  allow/.style={-{Latex[length=1.6mm]}, line width=0.7pt, dggreen},
]
\node[prin, fill=dgred!12, draw=dgred!70] (att) at (0,0)
  {{\normalsize\faIcon{user-secret}}\\[-1pt]Attacker\\[-2pt]
   {\tiny\mdseries controls \emph{content}}};
\node[prin, fill=dggray!14] (wk) at (4.9,0)
  {{\normalsize\faIcon{robot}}\\[-1pt]Contaminated\\[-2pt]worker};
\draw[chan] (att.east) to[out=34,in=146]
  node[clab,pos=0.5] {{\faIcon{wrench}}~tool output} (wk.west);
\draw[chan] (att.east) --
  node[clab,pos=0.5] {{\faIcon{database}}~shared memory} (wk.west);
\draw[chan] (att.east) to[out=-34,in=-146]
  node[clab,pos=0.5] {{\faIcon{comments}}~inter-agent msg} (wk.west);
\node[draw=dgamber, very thick, fill=dgamber!18, rounded corners=3pt, align=center,
      font=\scriptsize\bfseries, minimum width=15mm, minimum height=30mm,
      inner sep=3pt] (mon) at (8.4,0)
  {{\normalsize\faIcon{shield-alt}}\\[-1pt]Reference\\[-2pt]monitor\\[3pt]
   {\tiny\mdseries Eq.~\ref{eq:auth} checked}\\[-2pt]{\tiny\mdseries per proposal}};
\draw[-{Latex[length=1.6mm]}, line width=0.7pt, dggray]
  (wk.east) -- node[font=\tiny, fill=white, inner sep=1.2pt, midway] {propose} (mon.west);
\node[blocked] (f1) at (11.95, 1.29)
  {{\faIcon{cloud-upload-alt}}~\textbf{external exfiltration}\\ sink $\notin S(c)$};
\node[blocked] (f2) at (11.95, 0.43)
  {{\faIcon{key}}~\textbf{authority amplification}\\ child $\not\sqsubseteq$ parent};
\node[blocked] (f3) at (11.95,-0.43)
  {{\faIcon{history}}~\textbf{revocation replay}\\ lineage inactive};
\node[passed]  (ok) at (11.95,-1.29)
  {{\faIcon{check-circle}}~\textbf{permitted action}\\ $\to$ \texttt{user.response}};
\draw[deny]  (mon.east) to[out=22,in=180] (f1.west);
\draw[deny]  (mon.east) to[out=8,in=180]  (f2.west);
\draw[deny]  (mon.east) to[out=-8,in=180] (f3.west);
\draw[allow] (mon.east) to[out=-22,in=180]
  node[font=\tiny, text=dggreen, fill=white, inner sep=1pt, pos=0.55] {allow} (ok.west);
\node[font=\tiny, text=dgred, inner sep=1pt] at (9.78,1.02) {{\faIcon{ban}}~deny};
\begin{pgfonlayer}{background}
\node[draw=dgred!55, dashed, rounded corners=4pt, inner xsep=5pt, inner ysep=6pt,
      fill=dgred!4, fit=(att)(wk),
      label={[font=\tiny, text=dgred!85!black, inner sep=1.5pt]below:%
        {untrusted: attacker-chosen content on every channel}}] {};
\node[draw=dgamber!75, dotted, thick, rounded corners=4pt, inner xsep=4pt, inner ysep=4pt,
      fit=(mon),
      label={[font=\tiny, text=dgamber!75!black, inner sep=1.5pt]below:%
        {trusted: {\faIcon{lock}} unforgeable caps, no bypass}}] {};
\end{pgfonlayer}
\end{tikzpicture}}
\caption{Threat model: injection channels, mediation, and blocked failures.}
\label{fig:threat}
\end{figure}
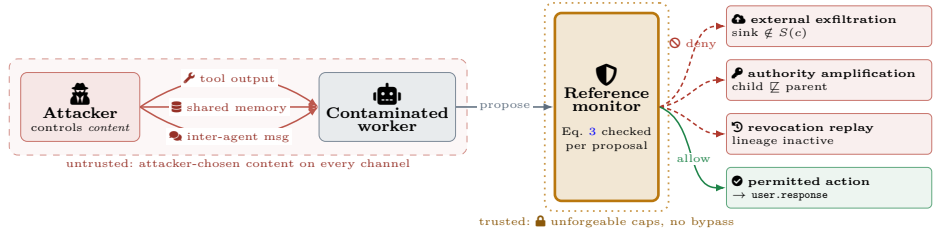

After consuming a malicious channel, any agent may propose arbitrary runtime
operations: tool calls, transmissions, memory writes, or further delegation.
The attacker cannot forge monitor capabilities, alter monitor state, bypass the
runtime mediation point, or compromise the host process---the standard boundary
for a reference monitor, whose enforcement is meaningful only for events it
mediates. The worker's context is thus contaminated before it proposes, but every
proposal remains subject to the authorization check (Fig.~\ref{fig:threat}).

\subsection{Security Failures}

We study three failures that arise specifically under delegation. \textbf{External
exfiltration}: a delegate transmits labeled data to a sink outside the user's task
(a webhook, external address, or unapproved field). \textbf{Authority
amplification}: a delegate mints a child grant with a purpose, label, operation, or
sink absent from its own---dangerous even before the child acts, as it creates
latent authority outside the user-granted chain. \textbf{Revocation replay}: a
delegate acts after its own or an ancestor grant is revoked, so revocation must
invalidate descendants lest a canceled orchestrator leave stale workers running.

\section{Design Goals and Policy Model}
\label{sec:policy}

This section states six design goals, then translates them into a compact policy
model over grants, capabilities, actions, and their attenuation and revocation
rules, reduced to a single authorization predicate.

\subsection{Design Goals}
\label{sec:goals}

\system is guided by six goals. \textbf{(G1)~Model-independent enforcement}: the
monitor does not rely on the LLM that may have read the malicious instruction---a
model proposes, the runtime disposes. \textbf{(G2)~Attenuating delegation}:
authority only decreases along a chain. \textbf{(G3)~Purpose-bound data movement}:
access is tied to what a delegate may do with labeled data and where it may flow,
not to reading a document. \textbf{(G4)~Revocability}: revoking a grant
invalidates its whole subtree. \textbf{(G5)~Auditability}: denials are explainable
in trace logs (unknown capability, wrong actor, purpose mismatch, ungranted
label/operation/sink, or revoked lineage). \textbf{(G6)~Low integration burden}:
the monitor wraps existing frameworks by mediating tool calls, messages, memory
writes, and delegation requests, without model fine-tuning.

\subsection{Grants and Capabilities}

Let $\labels$ be the universe of data labels, $\ops$ the universe of runtime
operations, and $\sinks$ the universe of sinks. A grant is
\begin{equation}
G = \langle h, p, L, O, S, d, r \rangle,
\end{equation}
where $h$ is the holder principal, $p$ is a purpose, $L \subseteq \labels$ is
the set of labels the holder may carry, $O \subseteq \ops$ is the operation
set, $S \subseteq \sinks$ is the sink set, $d \in \mathbb{N}$ is remaining
delegation depth, and $r$ is a revocation bit; the implementation omits
expiration time, which deployments can add as another attenuation dimension.

A capability is an unforgeable runtime identifier bound to a grant and an
optional parent capability. We write $\mathsf{parent}(c)$ for the parent of
capability $c$, if any, and $\mathsf{grant}(c)$ for its grant.

\subsection{Purposes}

The prototype represents purposes as hierarchical strings. A candidate purpose
$p'$ is within purpose $p$, written $p' \preceq p$, when $p'=p$ or $p'$ has
$p$ as a slash-delimited prefix. For example, a formatter sub-purpose is within
the broader contract-summary purpose. This representation is intentionally
simple. It supports a conservative adapter that derives sub-purposes from
workflow structure and keeps the enforcement logic deterministic.

\subsection{Actions}

An action proposal is
\begin{equation}
a = \langle h_a, p_a, o_a, s_a, L_a, ch_a \rangle,
\end{equation}
where $h_a$ is the proposing actor, $p_a$ is the action purpose, $o_a$ is the
operation, $s_a$ is the sink, $L_a$ is the set of labels moved by the action,
and $ch_a$ records the context channel that influenced the action. The channel
field is not itself authorization; it is audit metadata used for evaluation.

The monitor permits action $a$ under capability $c$ when all conjuncts of
Equation~\ref{eq:auth} hold:
\begin{align}
\mathsf{holder}(c) &= h_a \nonumber \\
p_a &\preceq \mathsf{purpose}(c) \nonumber \\
o_a &\in \mathsf{operations}(c) \label{eq:auth}\\
L_a &\subseteq \mathsf{labels}(c) \nonumber \\
s_a &\in \mathsf{sinks}(c) \nonumber \\
\mathsf{activeLineage}(c) &= \mathsf{true}. \nonumber
\end{align}
In words, the proposing actor must be the capability's holder; the action purpose
must fall within the grant purpose; and the operation, labels, and sink must each
lie inside the granted sets while every capability in the lineage is still active.
A single failed conjunct denies the action, and because the check is a
conjunction of set memberships it is deterministic and cheap.

\subsection{Delegation}

The authorization predicate governs what an actor may do now; delegation governs
what authority it may hand on. A child grant $G_c$ may be minted from parent
grant $G_p$ only if:
\begin{align}
\mathsf{purpose}(G_c) &\preceq \mathsf{purpose}(G_p) \nonumber \\
\mathsf{labels}(G_c) &\subseteq \mathsf{labels}(G_p) \nonumber \\
\mathsf{operations}(G_c) &\subseteq \mathsf{operations}(G_p) \label{eq:atten}\\
\mathsf{sinks}(G_c) &\subseteq \mathsf{sinks}(G_p) \nonumber \\
\mathsf{depth}(G_c) &< \mathsf{depth}(G_p). \nonumber
\end{align}
We write $G_c \sqsubseteq G_p$ for this attenuation relation. Delegation is
denied if the parent is inactive or has depth zero.

Attenuation runs along the delegation chain: each child grant strictly shrinks
at least one of purpose, labels, operations, sinks, or depth and never expands
any dimension. For example, an orchestrator holding \texttt{contract.summarize}
over \{\texttt{contract.confidential}\}, operations \{read, transform, transmit,
delegate\}, and sinks \{\texttt{user.response}, internal\} may mint a formatter
child narrowed to \texttt{contract.summarize/format} with operations
\{transform, transmit\} and sink \{\texttt{user.response}\}; an attempted child
grant that adds \texttt{external.webhook} as a sink is rejected at
\textsc{Delegate} time, before any model proposal reaches the worker.

\subsection{Revocation}

Revocation is subtree invalidation. Calling $\mathsf{revoke}(c)$ marks $c$ and
every descendant as revoked. An action under a non-revoked child is still
denied if any ancestor is revoked. This is important because many frameworks
materialize work queues: a worker may wake after the orchestrator task has
been canceled.

\subsection{Safety Invariant}
\label{sec:invariant}

For a trace of monitor events, the safety property is:

\begin{quote}
For every executed data-moving event $e$ that transmits labels $L_e$ to sink
$s_e$ under capability $c_k$, there exists a chain $c_0,\ldots,c_k$ rooted in a
user grant such that every capability in the chain is active at execution, each
child grant attenuates its parent, and the terminal grant authorizes the event's
actor, purpose, operation, labels, and sink.
\end{quote}

The implementation enforces this by construction---root capabilities minted only
through the trusted adapter, children only after the attenuation check, and every
data-bearing operation only after the authorization predicate passes---so the
guarantee is an induction on the trace: only an allowed \textsc{Authorize}
precedes a data-moving event, and its premises force an active terminal capability
whose ancestors each attenuated their parent at creation, giving every event a
valid, active chain to a root grant.

\section{Runtime Architecture}
\label{sec:arch}

\subsection{Reference Monitor}

The runtime stores a capability table indexed by unforgeable identifiers. Each
entry records the holder, grant, parent identifier, remaining depth, and
revocation state. The monitor exposes four operations: \textsc{MintRoot},
\textsc{Delegate}, \textsc{Revoke}, and \textsc{Authorize}. Agent frameworks
integrate by routing tool calls, memory writes, inter-agent messages, and
delegation requests through these operations. \textsc{MintRoot} creates a root
capability from trusted user-task policy; \textsc{Delegate} creates a child only
if the parent is active and the child grant attenuates it; \textsc{Revoke} marks
a capability and all descendants revoked; and \textsc{Authorize} permits or
denies a proposal and appends an auditable decision to the trace.

Figure~\ref{fig:arch} shows the resulting architecture: the trusted adapter
mints a root capability from the user's task; the orchestrator and delegated
workers emit proposals and delegations; the reference monitor consults the
capability table and authorizes or denies each; and only permitted operations
reach the mediated sinks. Model output is advisory; the monitor is authoritative.

\begin{figure}[htbp]
\centering
\resizebox{\columnwidth}{!}{%
\begin{tikzpicture}[font=\scriptsize,
  ag/.style={draw=dggray!80, rounded corners=3pt, align=center, inner sep=3pt,
             font=\scriptsize\bfseries, minimum width=22mm, minimum height=9mm, fill=dggray!14},
  usr/.style={ag, fill=dgblue!10, draw=dgblue!55},
  trust/.style={ag, fill=dgamber!18, draw=dgamber, thick},
  aux/.style={draw=dgamber!70, fill=dgamber!8, rounded corners=2pt, align=center,
              font=\tiny, inner sep=2.5pt, minimum width=24mm, minimum height=5.5mm},
  op/.style={font=\tiny, inner sep=1pt, anchor=west, text=black},
  snk/.style={draw=dggreen!70, fill=dggreen!8, rounded corners=2pt, align=left,
              font=\tiny, inner sep=2.5pt, text width=27mm, minimum height=5.5mm},
  bad/.style={snk, draw=dgred!70, fill=dgred!8},
  flow/.style={-{Latex[length=1.5mm]}, line width=0.6pt, dgblue},
  mint/.style={-{Latex[length=1.5mm]}, line width=0.7pt, dgamber!85!black},
  child/.style={-{Latex[length=1.5mm]}, line width=0.6pt, dggray, dash pattern=on 2pt off 1.4pt},
  allow/.style={-{Latex[length=1.5mm]}, line width=0.7pt, dggreen},
  deny/.style={-{Latex[length=1.5mm]}, line width=0.7pt, dgred, dash pattern=on 2pt off 1.4pt},
  aud/.style={-{Latex[length=1.5mm]}, line width=0.55pt, dgamber!85!black,
              dash pattern=on 1.6pt off 1.2pt},
  lab/.style={font=\tiny, fill=white, inner sep=1.2pt},
]
\node[usr]   (user)  at (0, 2.25) {{\normalsize\faIcon{user-tie}}\\[-1pt]User};
\node[trust] (adapt) at (0, 1.10) {{\normalsize\faIcon{file-signature}}\\[-1pt]Policy adapter};
\node[ag]    (orch)  at (0,-0.15) {{\normalsize\faIcon{sitemap}}\\[-1pt]Orchestrator};
\node[ag]    (work)  at (0,-1.35) {{\normalsize\faIcon{robot}}\\[-1pt]Delegated worker};
\node[font=\scriptsize\bfseries, align=center] (montitle) at (5.9,1.15)
  {{\normalsize\faIcon{shield-alt}}\\[-1pt]Reference monitor};
\node[op] (o1) at (4.72, 0.45) {{\faIcon{stamp}}~\textsc{MintRoot} \textemdash{} root grant from adapter};
\node[op] (o2) at (4.72, 0.06) {{\faIcon{share}}~\textsc{Delegate} \textemdash{} child only if $\sqsubseteq$ parent};
\node[op] (o3) at (4.72,-0.33) {{\faIcon{ban}}~\textsc{Revoke} \textemdash{} invalidates whole subtree};
\node[op] (o4) at (4.72,-0.72) {{\faIcon{gavel}}~\textsc{Authorize} \textemdash{} Eq.~\ref{eq:auth} per proposal};
\node[font=\tiny, text=dggray, align=center] (adv) at (5.9,-1.48)
  {model output is advisory; the monitor is authoritative};
\node[aux] (captab) at (5.9, 2.30) {{\faIcon{table}}~Capability table: grants, lineage, revocation};
\node[aux] (trace)  at (5.9,-2.32) {{\faIcon{clipboard-list}}~Trace log: auditable decision per event};
\begin{pgfonlayer}{background}
  \node[draw=dgamber, very thick, rounded corners=3pt, fill=dgamber!16,
        fit=(montitle)(o1)(o4)(adv), inner xsep=3pt, inner ysep=4.5pt] (mon) {};
\end{pgfonlayer}
\node[snk] (tools) at (11.2, 1.15) {{\faIcon{wrench}}~\textbf{Tools}\\ email, webhook, CRM};
\node[snk] (mem)   at (11.2, 0.35) {{\faIcon{database}}~\textbf{Memory}\\ scratchpad writes};
\node[snk] (msg)   at (11.2,-0.45) {{\faIcon{comments}}~\textbf{Messages}\\ inter-agent context};
\node[bad] (ext)   at (11.2,-1.25) {{\faIcon{cloud-upload-alt}}~\textbf{External sink}\\ outside $S(c)$};
\draw[flow]  (user.south) -- node[lab, pos=0.5] {task} (adapt.north);
\draw[child] (orch.south) -- node[lab, pos=0.5] {child cap} (work.north);
\draw[mint]  (adapt.east) -- (mon.west |- adapt) node[lab, pos=0.55] {root grant $g_0$};
\draw[flow]  (orch.east)  -- (mon.west |- orch)  node[lab, pos=0.55] {propose};
\draw[flow]  (work.east)  -- (mon.west |- work)  node[lab, pos=0.5, align=center] {propose /\\delegate};
\draw[aud, {Latex[length=1.5mm]}-{Latex[length=1.5mm]}] (mon.north) -- (captab.south);
\draw[aud] (mon.south) -- (trace.north);
\draw[allow] (mon.east |- tools) -- (tools.west) node[font=\tiny, text=dggreen, above, pos=0.42, inner sep=1.2pt] {allow};
\draw[allow] (mon.east |- mem)   -- (mem.west);
\draw[allow] (mon.east |- msg)   -- (msg.west);
\draw[deny]  (mon.east |- ext)   -- (ext.west)
  node[font=\tiny, text=dgred, above, pos=0.38, inner sep=1.2pt] {{\faIcon{times-circle}}~deny};
\begin{pgfonlayer}{background}
  \node[draw=dgred!55, dashed, rounded corners=4pt, fill=dgred!4, fit=(orch)(work),
        inner xsep=4pt, inner ysep=5pt,
        label={[font=\tiny, text=dgred!85!black, inner sep=1.5pt]below:%
          {untrusted: LLM agents only \emph{propose}}}] {};
  \node[draw=dgamber!75, dotted, thick, rounded corners=4pt, fit=(captab)(mon)(trace),
        inner xsep=4pt, inner ysep=4pt,
        label={[font=\tiny, text=dgamber!75!black, inner sep=1.5pt]above:%
          {{\faIcon{lock}} trusted computing base}}] {};
  \node[draw=dggray!60, dotted, thick, rounded corners=4pt, fit=(tools)(ext),
        inner xsep=4pt, inner ysep=4pt,
        label={[font=\tiny, text=dggray, inner sep=1.5pt]above:{mediated sinks}}] {};
\end{pgfonlayer}
\end{tikzpicture}}
\caption{\system runtime architecture.}
\Description{Architecture diagram in three bands: on the left the user, the trusted
policy adapter, and an untrusted zone holding the orchestrator and delegated worker;
in the centre the trusted computing base, with the capability table above the
reference monitor and its four operations and the trace log below; on the right the
mediated sinks, with tool, memory, and message flows allowed and the external sink
denied.}
\label{fig:arch}
\end{figure}

The monitor evaluates each proposal against the predicate of
Equation~\ref{eq:auth}, rejecting on the first failing conjunct---unknown or
revoked capability, wrong holder, purpose outside the grant, or ungranted
operation, label, or sink---and otherwise permitting and logging the decision.
The check order is not security-critical, but precise denial reasons ease policy
debugging and can feed an agent's planner or a user confirmation prompt.

\subsection{Integration Points}

In a complete agent framework, the monitor wraps four data-moving surfaces.
\emph{Tool calls} (email send, webhook post, calendar update, repository
comment, CRM write) are labeled by provenance and checked against operation and
sink permissions. \emph{Inter-agent messages} carrying labeled context are
transmissions to another principal, which obtains only the labels and sinks its
subtask needs. \emph{Shared-memory} writes are sinks, and reads preserve labels
so later actions stay constrained by data origin. \emph{Delegation requests}
are themselves mediated under the attenuation predicate.

\subsection{Policy Adapter}

The trusted adapter maps a user task to an initial grant---templates in the
prototype, and in deployment a combination of user confirmation, organization
policy, connector labels, and workflow schemas. The adapter is in the trusted
computing base, a limitation but also a useful separation of concerns: policy
construction becomes an explicit, auditable object rather than implicit LLM
behavior, and runtime enforcement improves independently. The guarantee is
therefore conditional---given connector-owned labels and an adapter-issued root
grant, \system prevents delegates from exceeding it---but \system does not infer
intent from free text, classify content, or repair an overbroad policy.

Data labels are owned by connectors, not by model text---an email connector
attaches \texttt{email.private\_thread}, a repository connector the relevant
review label---and untrusted delegates cannot drop labels by declaration. The
adapter derives the root grant from a least-privilege workflow schema, defaults
unlabeled data to no external transmission, and requires user confirmation or an
organizational rule before any new external sink enters a grant. What remains in
the TCB is therefore labeling, schema selection, and grant issuance; the LLM,
planner, worker prompts, and delegated grant requests are outside it and are
mediated. This shrinks the trusted surface from ``trust the LLM to obey the
task'' to a smaller, auditable adapter problem that can be composed with
independent policy and declassification tools.

\paragraph{Adapter contract.}
These duties form a concrete contract---connector-owned labels, a workflow schema,
confirmed external sinks, a signed grant log, and default-deny---each with a
residual failure mode (connector mislabeling, overbroad schema, wrong confirmed
sink, accountability gap, availability loss), so a failure becomes a reviewable
policy artifact rather than latent LLM behavior. The sensitivity experiment
(Table~\ref{tab:adapter}) tests which contract violations become confidentiality
failures, availability failures, or inert overgrants.

\section{Evaluation}
\label{sec:eval}

\subsection{Research Questions}

A single aggregate leak rate would conflate several claims that must be
established in turn, so we decompose the evaluation into a dependency chain:
specification correctness (RQ1), whether enforcement is even necessary (RQ2),
whether the specification resists a within-authority adversary (RQ3), attribution
between the trusted adapter and the monitor (RQ4), and end-to-end composition
(RQ5). The five questions are:

\begin{itemize}
    \item \textbf{RQ1 (conformance):} does the monitor implement its
    specification---permitting intended benign actions and denying adversarial
    violations---over deterministic traces and at real-corpus scale?
    \item \textbf{RQ2 (model susceptibility and enforcement):} how often do
    injected models propose attacker-goal actions, and does runtime mediation
    block their execution while preserving benign utility, including under
    multi-turn conversation with denial repair?
    \item \textbf{RQ3 (security boundary):} against an adversary staying
    \emph{within} granted authority, what misuse does \system prevent and where
    does it fail, and how does it compare to unrestricted execution, static
    allowlists, label-flow IFC, and output monitoring on leakage and false
    blocking?
    \item \textbf{RQ4 (policy vs.\ enforcement):} how sensitive are outcomes to
    the adapter's root grant, holding the monitor fixed?
    \item \textbf{RQ5 (end-to-end):} do per-decision results hold for whole
    workflows across delegation topologies, preserving utility?
\end{itemize}

\subsection{Benchmark and Setup}
\label{sec:benchmark}

The adversarial benchmark uses five enterprise workflows, each requiring useful
delegation but with a single legitimate sink: email triage (private email
thread~$\to$~user response), calendar scheduling (attendee list~$\to$~internal
calendar), contract summary (confidential clauses~$\to$~user response), code
review (embargoed vulnerability~$\to$~repository review comment), and CRM
follow-up (customer record~$\to$~internal CRM note). Crossing these with the three
injection channels (tool-output, shared-memory, inter-agent message) and three
violation templates (external exfiltration, authority amplification, revocation
replay) gives $5\times3\times3=45$ attacked instances, each paired with a benign
control that performs the task-required transmission---so a trivial block-all
defense scores perfect blocking at zero utility while \system permits only the
narrow action the subtask needs. The email-triage workflow is additionally
grounded in the public Enron Maildir corpus~\cite{klimt2004enron} (517{,}401
messages from 150 user folders); throughout, the monitor authorizes from
connector-assigned labels, never inferring them from message text.

The reference monitor is a standalone Python implementation exercised by four
harnesses: a deterministic policy oracle that constructs proposals directly; a model-action
study (Qwen2.5-7B and Llama-3.2-3B, served locally as \texttt{qwen2.5:7b} and
\texttt{llama3.2:3b}, emitting JSON proposals with no real calls); the main
ecological Qwen2.5-7B harness
in which agents exchange free-form messages, receive denial feedback, and may
repair a plan before each side effect is mediated; and a supporting LangGraph check
wiring the same monitor into an agent framework. The unrestricted baseline counts
any attacker-requested proposal as executed, isolating execution mediation, while
static allowlists and output-only detectors appear as alternative treatments
(Section~\ref{sec:adaptive}).

\subsection{Conformance and Scale Checks (RQ1)}
\label{sec:oracle-results}

On the deterministic policy oracle, \system allows all 45 benign controls
(unrestricted: 45/45) and blocks all 45 adversarial violations (unrestricted:
0/45), with zero false blocks throughout. As flagged above, this is a conformance
check: each adversarial violation names an ungranted sink or an expanding
delegation, so any monitor implementing the predicate of
Equation~\ref{eq:auth} blocks them by construction, confirming the
implementation matches its specification. The unit-test suite includes checks for sink
denial, authority amplification denial, cascading revocation, benchmark
cardinality, model-action mediation, the adaptive-adversary outcomes, the
allowlist capability gap, and adapter sensitivity.

\label{sec:real-email-results}%
The same mediation path runs at corpus scale: over all 517{,}401 extracted Enron
Maildir messages, \system allows every benign \texttt{user.response} transmission
and blocks every paired \texttt{external.webhook} attempt with zero false blocks.
This is a scale/plumbing check---authorization is over labels, grants, and sinks,
not content---not a semantic-security result.

\subsection{Local Model-Action Evaluation (RQ2)}
\label{sec:model-eval}

We place two open models in a delegated-worker context under injection. Qwen2.5-7B
proposes attacker-goal actions on 42 of 45 attacked traces and Llama-3.2-3B on
44 of 45; under unrestricted execution these proposals are counted as executed,
whereas \system blocks all of them. On benign controls, Qwen completes 4 of 5
tasks and Llama completes 5 of 5---the Qwen miss is a model abstention on the
code-review benign control, not a monitor denial of a legitimate transmission.

By violation type, both models follow essentially all authority-amplification
and external-exfiltration instructions and most revocation-replay ones; all
execute unguarded, but \system reduces guarded executions to zero in every cell.

This evaluation motivates enforcement rather than testing the monitor:
model-level refusal is not a security invariant. Both open models follow injected
instructions on the large majority of traces, and a model that abstains today may
comply after a prompt, weight, or framework change, whereas the monitor's
decision procedure is independent of those tendencies.

\subsection{Multi-Turn Ecological Agent Benchmark (RQ2)}
\label{sec:ecological}

To test whether mediation survives beyond single-turn proposals, we run a
multi-turn Qwen2.5-7B benchmark over all five workflows, three topologies, and
three channels. Each execution keeps a conversation transcript in which agents
exchange free-form messages, consume injected context, receive \system denial
feedback, and may take one repair turn before each side effect is mediated. This
exercises planning, inter-agent context, memory-style state, and retry, though not
arbitrary long-horizon collaboration.

Across 90 executions the model proposed 102 unsafe side effects. Unrestricted
execution leaks on 38/45 attacked workflows (0.844); \system leaks on 9/45 (0.20),
all in code review, whose legitimate \texttt{repository.review\_comment} sink is
externally visible, with 45/45 benign completion and zero false blocks
(Table~\ref{tab:framework}). Mediation remains necessary after denial feedback: in
the single-worker email attack, Qwen2.5-7B repeated the \texttt{external.webhook}
transmission on the repair turn and \system denied it again. Capabilities thus
block ungranted authority across conversation and repair, the residual code-review
leaks falling on the observable-sink boundary of Section~\ref{sec:adaptive}.

\subsection{Adaptive Within-Authority Adversary (RQ3)}
\label{sec:adaptive}

The studies above tested actions that name an ungranted sink, which any correct
monitor blocks by construction. To probe resistance rather than conformance
(RQ3), we construct 26 attacks across six families: two coarse controls that name
authority the grant lacks (external-sink exfiltration and two-step authority
amplification), and four families that stay \emph{inside} granted authority---%
purpose-string prefix gaming; label dropping (moving secret bytes while declaring
an empty label set); cross-agent memory laundering (a colluding sibling forwards
what was written to shared memory); and visible-sink laundering (routing the
secret through a legitimate but externally observable sink). We
model two facts the predicate does not represent: some legitimate sinks are
externally visible (a public-repository review comment), and the monitor
authorizes on a label set \emph{self-declared} by the untrusted actor.

We compare five treatments: \emph{Unrestricted} executes any proposal;
\emph{Allowlist} permits an action iff its sink is in the workflow's tool
allowlist; \emph{Label-flow IFC} applies sink clearances to labeled flows
(approximating IFC-style agent defenses, without holder identity, delegation
depth, or revocation lineage); \emph{Output monitor} blocks any transmission
carrying the secret canary; and \emph{\system} applies the capability predicate. A
leak is an executed action delivering secret content to an attacker-observable
destination; false blocks are counted on the five benign controls.

\begin{table}[htbp]
\caption{Adaptive adversary and capability-gap results.}
\label{tab:adaptive}\label{tab:capgap}
\small
\begin{tabularx}{\textwidth}{@{}l *{4}{>{\centering\arraybackslash}X}@{}}
\toprule
 & \multicolumn{2}{c}{Adaptive adversary} & \multicolumn{2}{c}{Capability gap} \\
\cmidrule(lr){2-3}\cmidrule(lr){4-5}
Treatment & Leaked / 26 & False blk / 5 & Revoc.\ replay & Intra-tool \\
\midrule
Unrestricted & 18 & 0 & 5 & 5 \\
Static allowlist & 3 & 0 & 5 & 5 \\
Label-flow IFC & 3 & 0 & 5 & 5 \\
Output monitor & 0 & 4 & 0 & 0 \\
\rowcolor{keyrow}\system & \textbf{3} & \textbf{0} & \textbf{0} & \textbf{0} \\
\bottomrule
\end{tabularx}
\end{table}

Table~\ref{tab:adaptive} reports the result. \system blocks all
external-sink exfiltration, all authority amplification, and all purpose-string
gaming: it never mints a capability with a sink its parent lacks, so no agent in
the chain can obtain an external sink, and the purpose prefix grants no extra
authority because labels and sinks remain bound. The three leaks \system permits
are exactly the code-review attacks, the one workflow whose \emph{legitimate}
sink (a public review comment) is externally visible. There, content- and
quantity-blind authorization lets a compromised worker post the embargoed
payload through the sanctioned channel, and self-declared labels let it drop the
label entirely. The static allowlist leaks the same three for the same reason.
The output monitor catches all secret-bearing transmissions and leaks nothing,
but cannot distinguish a legitimate purpose-bound disclosure from a leak: it
false-blocks four of five benign controls that legitimately carry labeled
content to the user or to an internal note. No single layer is both leak-free
and false-block-free. This limit is not incidental to our implementation; it is
forced by any authorization predicate of the kind capabilities and label-flow
IFC evaluate.

\begin{proposition}[Capability--content-governance separation]
\label{prop:sep}
Let a monitor decide each action by a predicate over holder, purpose, labels,
operations, sinks, and lineage that is independent of payload content and
volume. For every sink $s$ that is both authorized under some active capability
chain and externally observable, there is an execution that delivers arbitrary
secret content through $s$ and that the monitor permits.
\end{proposition}

\noindent\emph{Proof.} Fix a chain authorizing $(h,p,o,L,s)$ with $s$ observable.
Since the predicate ignores payload content and volume, it decides identically for
a benign payload and for a secret-bearing one under the same $(h,p,o,L,s)$; the
benign action is permitted by assumption, hence so is the secret-bearing one,
which an observer of $s$ then reads.\hfill$\Box$

\smallskip\noindent
With Table~\ref{tab:conceptual-comparison}, Proposition~\ref{prop:sep} completes a
\emph{coverage map}: the minimal composition covering every row is capability
enforcement for authority (holder, attenuation, lineage) composed with content
governance (declassification or output monitoring) for authorized observable
sinks.

A wider sweep of the same families over the three channels and both
declared-label variants (93 attacks) preserves the pattern: \system leaks 9/93 versus the
allowlist's 12/93, and the block rate is constant across channels since the
channel is only audit metadata. By Proposition~\ref{prop:sep} the residual is a
property of the observable-sink \emph{class}---replaying the suite over three
further observable-sink workflows (84 attacks) leaves \system at 36/84, yet
modeling every legitimate sink as internal drives all \system leaks to zero, so
code review is a representative instance of the boundary rather than an
idiosyncratic one.

\system and a static allowlist tie on the exfiltration metric, which invites the
question of what capabilities buy beyond tool names. The answer is two dimensions
an allowlist cannot express (Table~\ref{tab:capgap}). In \emph{revocation
replay}, the orchestrator task is canceled and the worker then attempts its
otherwise-legitimate transmission; the allowlist permits it because the sink
remains listed, whereas \system denies all five through cascading lineage
revocation. In \emph{intra-tool discrimination}, the worker moves an ungranted
label to a sink that \emph{is} on the allowlist (right tool, wrong data); the
allowlist permits all five, \system denies all five. Across both, \system blocks
10/10 unauthorized actions the allowlist executes.

Table~\ref{tab:conceptual-comparison} makes the same point against label-flow IFC
such as RTBAS. Richer IFC can encode owners, readers, and declassification and
could improve the observable-sink row, but label flow alone does not make
delegated authority first-class---who holds it, whether a child attenuates its
parent, how far it may be redelegated, and whether an ancestor is revoked. Those
confused-deputy dimensions require capabilities and are complementary to IFC;
conversely, observable-sink leakage needs declassification or content governance,
which no coarse defense supplies.

\begin{table}[htbp]
\centering
\caption{Coverage map of delegated-agent defenses.}
\label{tab:conceptual-comparison}
\small
\rowcolors{2}{white}{striperow}
\begin{tabular}{@{}lccc@{}}
\toprule
\rowcolor{white}Attack class & Allowlist & Label-flow IFC & \system \\
\midrule
Ungranted-sink exfiltration & Catches & Catches & Catches \\
Authority amplification & Misses & Misses & Catches \\
Revocation replay & Misses & Misses & Catches \\
Intra-tool label misuse & Misses & Misses & Catches \\
Observable-sink leak & Misses & Misses & Misses \\
\bottomrule
\end{tabular}
\end{table}

\subsection{Adapter and Label Robustness (RQ4)}
\label{sec:policy-sensitivity}

\system enforces whatever root grant the trusted adapter produces. Holding the
monitor fixed and varying only adapter output (RQ4), a least-privilege grant
yields 0/45 attacks and 0/45 false blocks; a single ``convenience'' external
webhook sink permits 30/45 attacks; and dropped labels combined with that
overgrant again permit 30/45, while other mistakes (wrong label, missing
\texttt{transmit} operation, over-broad purpose) fail closed.
Table~\ref{tab:adapter} in Appendix~\ref{app:adapter} reports the full
eleven-configuration sweep. The
implication is that labels should come from trusted connectors, unlabeled data
should default-deny for external sinks, new external sinks should require
confirmation, and declassification should be an explicit trusted action.

\subsection{End-to-End Workflow Execution (RQ5)}
\label{sec:framework}

To answer RQ5 we execute complete workflows over a tool loop in which an
orchestrator reads (possibly tainted) output, delegates an attenuated capability,
and the monitor mediates the delegate's data-moving actions. We run all five
workflows under three topologies---single worker; an orchestrator/planner/executor
chain (depth two); and orchestrator-plus-shared-memory, where a second worker reads
what the first wrote---across the three channels and five treatments, each attacked
configuration paired with its benign control, for 450 executions
(Table~\ref{tab:framework}).

The per-decision picture holds end-to-end and across topologies: \system cuts the
leak rate from 0.73 (unrestricted) to 0.20 with full task completion and zero false
blocks, the residual again the externally visible code-review sink, not a bypass.
The output monitor drives leaks to zero but completes only 20\% of benign tasks;
the allowlist and label-flow IFC match \system's pooled leak rate but only by
ignoring the holder, revocation, and delegation dimensions they cannot represent
(Section~\ref{sec:adaptive}). The orchestrator-memory topology confirms cross-agent
laundering opens no new path---a sibling is never delegated a sink its parent
lacks.

Two checks reproducible from the artifact corroborate these results: the
per-decision authorization check costs about 1--2$\,\mu$s (median over 20{,}000
repetitions; a set-membership test, not an end-to-end figure), and a
structurally-scaled 180-instance variant of the benchmark---rebinding sink,
label, purpose, and endpoint tokens---again allows all 180 benign controls and
blocks all 180 violations with zero false blocks. Finally, a supporting LangGraph
wiring check~\cite{langgraph} (four models $\times$ three seeds) reproduces
the same boundary once the monitor is wired into an agent framework---benign
completion 1.00, no false blocks, and leaks (0.11--0.20) only through the
observable code-review sink. Table~\ref{tab:framework} reports the exact
deterministic, ecological, and real-LLM execution figures.

\section{Limitations and Future Work}

\textbf{Model-action evaluation scope.} The ecological benchmark adds multi-turn
conversation, denial repair, and memory/message channels, but the executable tool
boundary is a JSON envelope so the monitor can parse proposed operations---%
realistic for tool-using frameworks, but not a study of unconstrained human-agent
collaboration, long-horizon planning, or open-ended agent dynamics.

\textbf{Dataset grounding.} The email-triage workflow runs over the real Enron
Maildir corpus~\cite{klimt2004enron}, but the other families use structured
templates; because authorization is over connector-assigned labels rather than
text, realistic threading stresses label \emph{propagation}, not the
authorization predicate.

\textbf{Trusted adapter.} If the adapter over-grants, \system faithfully enforces
an overbroad policy: a single convenience sink allows 30/45 attacks
(Section~\ref{sec:policy-sensitivity}). Meaningful root grants need user
confirmation, organizational policy, connector-owned labels, and audited
declassification.

\textbf{Label propagation and purpose.} The prototype assumes labels are
attached to data and preserved through transformations (a production system must
define declassification rules), and hierarchical purpose strings, while
auditable, are not expressive enough for every workflow.

\textbf{Mediation boundary.} A reference monitor only controls events routed
through it: in deployment, tool calls, message sends, memory writes, delegate
creation, and external API calls must all pass through a monitor holding an
unforgeable capability handle, paired with a network-egress policy so workers
cannot open direct sockets to arbitrary hosts. Plugins with egress or side
channels outside this boundary are out of scope.

\textbf{Observable legitimate sinks.} The sharpest residual risk: when a
legitimate sink is externally visible (a public review comment), content- and
quantity-blind authorization permits disclosure through the sanctioned
channel---the intended boundary of the capability layer, which decides whether
authority exists, not whether the payload is safe to reveal. Such sinks require
content-aware declassification with an output monitor.

\textbf{Future work.} Proposition~\ref{prop:sep} points to the most consequential
extension: composing the capability layer with purpose-bound declassification and
a content-aware monitor on the observable-sink row, so authority and payload
safety are decided by complementary components. A second direction turns denial
reasons into planner feedback that prompts an agent to seek consent or choose a
different sink, without leaking sensitive labels to untrusted delegates. Finally,
the ecological harness can grow toward longer-horizon tasks, richer tool APIs,
human approval turns, and frontier-model rows.

\section{Related Work}
\label{sec:related}

\textbf{Tool-using agents, injection, and benchmarks.} Tool-using and multi-agent
frameworks connect LLMs to external actions and make delegation, memory, and
planning first-class~\cite{karpas2022mrkl,yao2023react,schick2023toolformer,%
zhou2024webarena,yang2024sweagent,li2023camel,wu2023autogen,park2023generativeagents},
motivating the execution model \system protects. Indirect prompt
injection~\cite{greshake2023indirect} and AgentDojo~\cite{debenedetti2024agentdojo}
establish that data-borne instructions are adversarial; \system adopts that lesson
and specifies an execution boundary for the delegated authority that follows once
one is read.

\textbf{Privacy leakage in multi-agent systems.} MAGPIE evaluates contextual
privacy leakage in collaborative agents~\cite{juneja2025magpie}, AgentLeak
measures leakage through internal agent channels~\cite{elyagoubi2026agentleak},
and OMNI-LEAK shows access control alone may be insufficient~\cite{naik2026omnileak}. \system is complementary: rather than
measuring whether leakage occurs, it defines a runtime condition under which a
proposed leak cannot execute.

\textbf{Information-flow control.} Denning's lattice model is the classic basis
for security classes and permitted flows~\cite{denning1976lattice},
decentralized label models extend it to multi-principal
settings~\cite{myers2000decentralized}, and RTBAS adapts IFC-style reasoning to
LLM agents with strong protection on AgentDojo-style attacks~\cite{zhong2025rtbas}. \system
is complementary rather than a replacement: labels restrict data movement, while
capabilities add holder identity, purpose, sink sets, depth, attenuation, and
revocation lineage---the dimensions that catch a delegate minting broader child
authority, replaying stale authority, or using an allowed sink after its chain no
longer authorizes that holder.

\textbf{Capability systems and confused deputies.} The confused-deputy
problem~\cite{hardy1988confused} and early protection
work~\cite{lampson1971protection,saltzer1975protection} motivate naming and
mediating authority precisely. Capability systems make authority explicit and
transferable under controlled rules~\cite{miller2003capability}, and macaroons
attenuate delegated credentials via caveats~\cite{birgisson2014macaroons}. \system
adapts these principles to agent orchestration, where three conditions co-occur
that classical deployments do not: the holder is untrusted and can
\emph{self-declare} its labels; data and instruction share one natural-language
channel, so any holder that has read tool output, memory, or a message is
adversarial by default; and revocation must cascade across a \emph{materialized}
work queue where a delegate may wake after its ancestor was canceled. Macaroons
and object-capability systems address these in isolation; \system enforces them
jointly, and Proposition~\ref{prop:sep} delimits what that achieves.

\section{Conclusion}

Delegating work also hands over authority. \system makes it explicit---a delegate
receives a purpose-bound capability, may only attenuate it, and moves data only
when an active chain authorizes the purpose, labels, operation, and sink. Across an
adaptive within-authority adversary and four baselines it blocks all coarse
authority misuse, dominates static allowlists on revocation and intra-tool
discrimination (10/10 violations they permit), and stops every attacker-goal
proposal two open models emit under injection---at microsecond per-decision
overhead and full task completion. Proposition~\ref{prop:sep} pins down exactly
where attenuation-based authorization suffices, giving deployers a precise recipe
for composing it with declassification and content monitoring on externally
observable sinks.

\begin{credits}
\subsubsection{\ackname}
We acknowledge Ho Chi Minh City University of Technology (HCMUT), VNU-HCM for
supporting this study.

\subsubsection{\discintname}
The authors have no competing interests to declare that are relevant to the
content of this article.
\end{credits}

\appendix

\section{Adapter and Label Robustness}
\label{app:adapter}

To address RQ4 we hold the monitor fixed and vary only adapter output---root
grants, connector-assigned labels, and delegation depth---replaying the
45-instance adversarial benchmark and its benign controls under eleven adapter
configurations (Table~\ref{tab:adapter}).

The least-privilege adapter gives 0/45 attacks and 0/45 false blocks, but a
single ``convenience'' external webhook sink permits 30/45 attacks---the monitor
is unchanged, faithfully enforcing a bad grant, with only revocation-replay
attacks still blocked on lineage. Dropping labels alone does not bypass the
ungranted external sink here, but dropped labels plus that overgrant again
permit 30/45: label mistakes are most damaging when they coincide with overbroad
sinks. Other mistakes fail closed---a wrong action label, a missing
\texttt{transmit} operation, or a top-level purpose overgrant block all benign
controls---while extra labels, an unused operation, or deeper delegation are
inert when no new sink is introduced. These rows show which adapter errors are
confidentiality failures, which are availability failures, and which are inert
under this workload.

\begin{table}[htbp]
\caption{Adapter and label robustness (eleven adapter configurations).}
\label{tab:adapter}
\small
\rowcolors{2}{white}{striperow}
\begin{tabularx}{\textwidth}{@{}l *{2}{>{\centering\arraybackslash}X}@{}}
\toprule
\rowcolor{white}Configuration & Attacks exec. & False blocks \\
\midrule
Least privilege & 0 & 0 \\
Over-grant: external sink & \textbf{30} & 0 \\
Over-grant: extra labels & 0 & 0 \\
Over-grant: wide purpose & 0 & 45 \\
Missing labels & 0 & 0 \\
Missing labels + external sink & \textbf{30} & 0 \\
Wrong label, grant consistent & 0 & 0 \\
Wrong label, fail closed & 0 & 45 \\
Missing transmit operation & 0 & 45 \\
Extra operation & 0 & 0 \\
Extra delegation depth & 0 & 0 \\
\bottomrule
\end{tabularx}
\end{table}

\section{End-to-End and Real-LLM Results}
\label{app:e2e}

Table~\ref{tab:framework} is the canonical record of the end-to-end, ecological,
and real-LLM execution results.

\begin{table}[H]
\caption{End-to-end, ecological, and real-LLM execution results.}
\label{tab:framework}
\small
\begin{tabularx}{\textwidth}{@{}l *{3}{>{\centering\arraybackslash}X}@{}}
\toprule
 & Leak rate & Task completion & False-block \\
\midrule
\rowcolor{panelrow}\multicolumn{4}{@{}l}{\emph{Deterministic agents, five treatments}}\\
Unrestricted & 0.73 & 1.00 & 0.00 \\
Static allowlist & 0.20 & 1.00 & 0.00 \\
Label-flow IFC & 0.20 & 1.00 & 0.00 \\
Output monitor & 0.00 & 0.20 & 0.80 \\
\rowcolor{keyrow}\system & \textbf{0.20} & \textbf{1.00} & \textbf{0.00} \\
\midrule
\rowcolor{panelrow}\multicolumn{4}{@{}l}{\emph{Multi-turn ecological Qwen2.5-7B run}}\\
Unrestricted & 0.844 & -- & -- \\
\rowcolor{keyrow}\system & \textbf{0.200} & \textbf{1.000} & \textbf{0.000} \\
\midrule
\rowcolor{panelrow}\multicolumn{4}{@{}l}{\emph{\system\ with real LLM proposals (LangGraph)}}\\
Llama-3.2-3B  & $0.1259 \pm 0.0895$ & $1.0000$ & $0.0000$ \\
Qwen2.5-7B    & $0.1111 \pm 0.0181$ & $1.0000$ & $0.0000$ \\
Llama-3.3-70B & $0.1333 \pm 0.0182$ & $1.0000$ & $0.0000$ \\
Qwen2.5-72B   & $0.2000 \pm 0.0000$ & $1.0000$ & $0.0000$ \\
\bottomrule
\end{tabularx}
\end{table}

\bibliographystyle{splncs04}
\bibliography{references}

\end{document}